\documentclass[12pt]{iopart}

\makeatother

\usepackage{amstext}
\usepackage{graphicx} 
\expandafter\let\csname equation*\endcsname\relax

\expandafter\let\csname endequation*\endcsname\relax
\usepackage{amsmath}

\usepackage[usenames,dvipsnames]{xcolor}
\usepackage{setspace}

\begin{document}

\title{Josephson coupling driven magnetoresistance in superconducting NiBi$_3$ nanowires}

\author{Laxmipriya Nanda$^1$, Bidyadhar Das$^1$, Subhashree Sahoo$^1$, Pratap K Sahoo$^{1, 2}$,  and Kartik Senapati$^{1, 2}$}
\address{$^1$School of Physical Sciences, National Institute of Science Education and Research,
\ An OCC of Homi Bhabha National Institute, Jatni-752050, Odisha, India.}
\address{$^2$Center for Interdisciplinary Sciences (CIS), NISER Bhubaneswar, Jatni-752050, Odisha, India.}
\ead{kartik@niser.ac.in}

\begin{abstract}
We present results of magnetoresistance (MR) measurements in granular NiBi$_3$ nanowires in the resistive state below the superconducting transition temperature. MR of 100 nm wide nanowires fabricated by focused Ion beam lithography from granular films of NiBi$_3$ with and without magnetic Ni impurity were compared. The nanowire containing high concentration of Ni impurity showed oscillations in MR and also exhibited a negative MR in certain temperature and field range. None of these effects were observed in the nanowire with no Ni impurities. Therefore, we argue that this effect is a result of the random Josephson couplings realized across superconducting NiBi$_3$ grains via magnetic inter grain regions. Such random couplings can cause local fluctuations in the density and sign of supercurrent, which can lead to negative MR and oscillations in MR, as proposed by Kivelson $\&$ Spivak [Kivelson et al. Phy. Rev. B. \textbf{45}, 10490 (1992)].
\end{abstract}

%
%
\submitto{\SUST}
%
\maketitle
%
%

\section*{Introduction}

Transport in granular superconducting films is often interpreted in terms of a coupled network of Josephson junctions, where individual grains with well defined superconducting order parameters couple across the grain boundaries. Abeles\cite{Abeles} has shown that in such an aggregate, the inter-grain charging energy arising from a finite inter-grain capacitance is of paramount importance for determining the critical response of the system. In multiply connected superconducting grains the distribution of inter-grain charging energy (equivalently, the distribution of capacitance) naturally brings in mesoscopic phase fluctuations into play. Since the phase and the number of Cooper pairs are conjugate variables\cite{anderson} inside a superconducting grain, phase fluctuations translate to Cooper pair number fluctuations in a reciprocal manner. In fact, Al'stuler and Spivak\cite{altshuler} have shown that in disordered SNS junctions, mesoscopic fluctuations in current can completely dominate the transport, especially in the limit of suppressed current across the junction. Such a situation can be realised when superconducting grains are separated by inter grain regions of magnetic nature\cite{altshuler}. Interestingly, the amplitude as well as the sign of the local order parameter can exhibit significant local variations inside the proximity coupled magnetic regions\cite{Ryazanov,Buzdin, robinson}. It has been demonstrated in hybrid ferromagnet-superconductor systems that the coupling between two superconducting banks can be $"Zero"$ or $"\pi"$ type depending on the thickness and magnetic exchange energy in the barrier\cite{Ryazanov,Buzdin, robinson,Robinson_Ni, ZeroPiTemp}. While the $"Zero"$ type coupling corresponds to a forward current the $"\pi"$ coupling corresponds to a reverse current across the barrier. In a random network of multiply connected superconducting grains one can expect a range of inter grain spacing leading to large fluctuations in the sign as well as amplitude of the local order parameters. Kivelson and Spivak \cite{Kivelson} have analyzed the consequence of fluctuations in the sign and amplitude of superfluid density on the average superconducting property of an ensemble of superconducting grains. They have shown that in such cases the magnetoresistnace can undergo sub-harmonic Aharanov-Bohm type oscillations and also, can result in negative magnetoresistance. Although their work was in the context of superconductor-to-insulator transition in granular films, the general outcomes of their calculations should hold, in general, if the assumptions of the model are satisfied in a system.

In order to test this hypothesis here we report the results of temperature dependent magnetoresistance measurements on a superconducting NiBi$_3$ nanowire fabricated by focused ion beam patterning from a granular NiBi$_3$ films. The choice of Superconducting NiBi$_3$ films was motivated by two factors. Firstly, our earlier work on this system\cite{Das2023} has shown that NiBi$_3$ films, prepared by co-evaporation of Ni and Bi metals, grow in a very granular form due to the large difference in the melting point of the two constituent metals. Secondly, co-evaporation of Ni and Bi leaves some Ni impurity in the system which can be varied, at least in a qualitative manner. Therefore, a granular superconductor with magnetic inter grain regions can be realized in this system, ideal for testing the model of Kivelson $\&$ Spivak\cite{Kivelson}. We find that low field magnetoresistance of a 100 nm wide NiBi$_3$ nanowire, with significant Ni impurity, exhibits MR oscillations and negative MR. A NiBi$_3$ nanowire of same width, albeit with no Ni impurity, did not show any of these effects. Comparing with the earlier reports on MR of superconducting nanowires, we argue that our results closely follow the general predictions of Kivelson$\&$ Spivak model\cite{Kivelson}. 

\section{Results}

\subsection{NiBi$_3$ films and fabrication of nanowires}
From our previous studies\cite{siva1, Siva2, bhatia2018superconductivity}, we have found that the intermetallic superconductor NiBi$_3$ with a T$_c \sim$4.2 K\cite{fujimori,silva,zhao}, forms naturally at the interface of Ni and Bi films by the process of reaction diffusion at room temperature. Due to the highly diffusive nature of this combination\cite{NiBi2,NiBi1,Siva2} it is possible to grow thin films of NiBi$_3$ by co-evaporation\cite{Das2023} of Ni and Bi. In this process, however, some surplus Ni and Bi impurities are expected to remain in the film. In order to control the magnetic Ni impurities, in this work, we have co-evaporated Ni and Bi through a magnetized mesh placed between the substrates and the sources. Further details are provided as supplementary information. In Fig. 1 (a) we compare the surface microstructures of $\sim$100 nm thick NiBi$_3$ films grown with and without a magnetic mesh, marked as "low-Ni" and "high-Ni" samples, respectively. Granular microstructure of the films is apparent in both the cases.
\begin{figure}[htbp!]
\centering
\includegraphics[width=12cm]{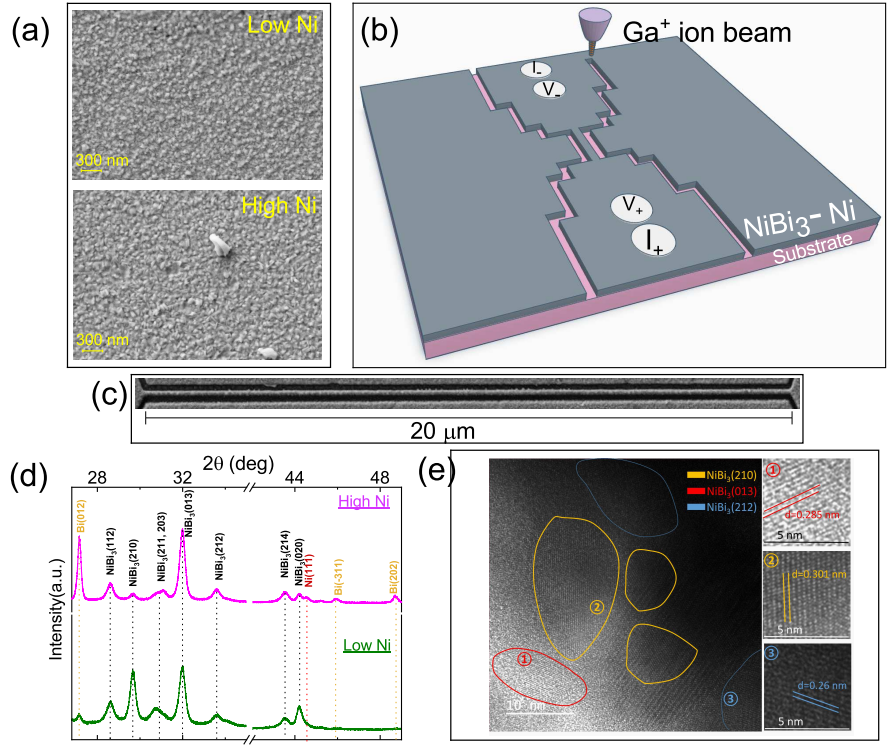}\\
	\caption{(a) Scanning electron micrograph of low-Ni and high-Ni films of NiBi$_3$ showing similar granular texture. (b) Schematic illustration of focused Ga ion beam patterning of the nanowire and contact pads from the film. The milling width was $\sim$200 nm. (c) A scanning electron micrograph of the 20 $\mu$m long and 100 nm wide nanowire. (d) Comparison of X-ray diffraction patterns of the high-Ni and low-Ni sample. For the low-Ni sample, which was deposited through a magnetized mesh, the Ni(111) peak has disappeared. (e) A representative high resolution transmission electron micrograph of the NiBi$_3$ film showing dispersion of randomly oriented grains. Some grains of different orientations have been marked in different colors. The magnified views of the fringe pattern for the numbered grains are also shown on 5 nm scale.}
\end{figure}
X-ray diffraction measurements of these two films (Fig. 1 (d)) reveals that the the strongest Ni (111) peak is almost non-existent in the low Ni film, grown with a magnetized mesh over the substrate. Henceforth we will use the nomenclature of "Low-Ni" and "High-Ni" to distinguish the two samples. 100 nm wide and 20 $\mu$m long nanowires were fabricated in a single-step milling procedure using 30keV focused Gallium ion beam which defined the nanowire and the contact pads in the same process. A schematic of the milling process is shown in the Fig. 1 (b). A scanning electron micrograph of the high-Ni nanowire is shown in Fig. 1 (c). The high resolution transmission electron micrograph (TEM) of the high-Ni sample, performed in a Jeol 200 keV system, is shown in Fig. 1 (e). Several grains oriented in different directions have been identified and shown in the magnified view. HRTEM images at another spot is given as supplementary Fig S1. The overall internal microstructure appears to be a distribution of randomly oriented NiBi$_3$ grains. Since the film has been prepared by co-evaporation the surplus Ni may be expected to be distributed uniformly throughout the material. With different amount of excess Nickel in the NiBi$_3$ film the magnetic nature of the inter grain region can change considerably, affecting the Josephson coupling.

\subsection{The resistive state of NiBi$_3$ nanowires below the superconducting transition temperature}
In this section we compare the transport behavior of the low-Ni and high-Ni nanowires below the superconducting transition temperature. Figure 2 (a) shows the superconducting transitions of the high-Ni sample in the film and nanowire forms. The resistances have been normalized at 5 K. The un-patterned film shows a sharp transition above 4.5 K, which is the best known transition temperature for NiBi$_3$ single crystals\cite{fujimori, silva, zhao}. The nanowire prepared from the same film shows a broader transition, below the initial transition of the electrodes close to 4.5 K. At the lowest temperatures the resistance of the nanowire flattens at a finite value. The same comparison of normalized resistance is shown in Fig. 2 (b) for the low-Ni sample in thin film and nanowire form. In this case also the un-patterned film shows a sharp transition around 4.5 K, while the nanowire resistance flattens at a non-zero value. The fact that the transitions of both the NiBi$_3$ films are sharp and same as single crystal T$_c$, implies that the surplus Ni in both cases are primarily in the inter-grain regions. Otherwise the T$_c$ of the individual grains, and consequently of the NiBi$_3$ films, would have been significantly suppressed by the pair breaking effects of the ferromagnetic Ni.

The resistive states of superconducting nanowires, below the transition temperature, have been a widely studied subject. The early pionieering work of Langer $\&$ Ambegaokar \cite{langer}, McCumber $\&$ Halperin\cite{mccumber}(LAMH theory) has established that thermal phase slip(TPS) events stabilize the resistive state in narrow superconductors near T$_c$. Although this random process is ubiquitous in all superconductors, the collective time averaged effect of these phase slip events become apparent in superconducting nanowires\cite{delacour2012quantum,zhao2016quantum}. It was later established \cite{giodano1,giodano2, bezryadin2000quantum,lau2001quantum,NbNNeg,giodano3,giodano4,altomare} that well below the transition temperature, quantum phase slips (QPS) can lead to a finite resistance in nanowire, typically below $\sim$0.7 T$_c$ \cite{giodano3,giodano1}. The combined effect of both mechanisms of phase slips leads to a finite resistance well below the transition temperature which increases with increasing temperature\cite{zhao2016quantum, bezryadin2008quantum}. 
\begin{figure}[t]
\centering
\includegraphics[width=12cm]{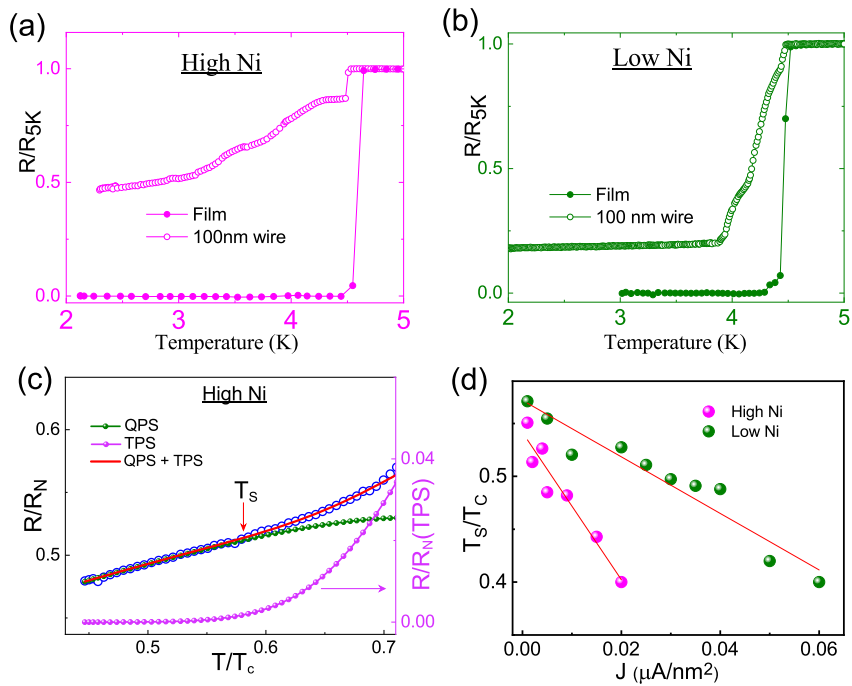}\\
	\caption{Panels (a) and (b) show the superconducting transitions of the NiBi$_3$ film and nanowire for the high-Ni and low-Ni samples, respectively. The resistances have been normalized to the resistance at 5K in both cases. While the films reach the zero resistance state, the nanowire remain in a phase slip driven resistive state down to the lowest temperature, consistent with reports on other superconducting nanowire systems. (c) A representative fit of the low temperature resistance (upto 0.7T$_c$) of the high-Ni NiBi$_3$ nanowire, measured with 40 $\mu$A current. The open symbols are the measured data and the solid line is the fit with the combined QPS and TPS models. The solid symbols plotted of the right hand axis and left hand axis are the TPS and QPS contributions to the fitting, respectively. The normalized temperature T$_s$ corresponds to the point upto which the QPS response dominates. The values of T$_s$ found from similar fittings of the low temperature resistances, measured at various currents, are plotted in the panel (d) for low-Ni and high-Ni nanowires. The solid line in panel (d) is only to guide the eye. In the high-Ni nanowire, QPS has a smaller temperature range of influence compared to the low-Ni nanowire.}
\end{figure}
We have fitted the low temperature resistances (upto 0.7 T$_c$) of the low-Ni and high-Ni nanowires to a combination of thermal and quantum phase slip models. Fig. 2 (c) shows a representative fit of the low temperature resistance of high-Ni nanowire measured with 40 $\mu$A current. The details of the fitting process are provided in supplementary information. In Fig. 2 (c) the open circles are the experimental points, the solid line is the overall fit, and the solid symbols corresponds to the QPS and TPS components of the fit. It shows that the QPS model fits very well upto a certain normalized temperature, T$_S$, above which the TPS component takes over. It has been shown earlier that the resistive state of the superconducting nanowires is current dependent\cite{giodano5,delacour2012quantum,zhao2016quantum}. Therefore we have performed resistance measurements of both nanowires at different bias current densities upto 0.06 $\mu$A/nm$^2$. Using the same fitting procedure we have estimated the temperature T$_S$ upto which the resistive state of the nanowires is QPS dominated. Fig 2 (d) shows the current dependence of T$_s$ for the low-Ni and high-Ni nanowires. In the high-Ni nanowire the QPS regime seems smaller, possibly due to the enhanced propensity of occurrence of thermal phase slip events in regions of partially suppressed order parameter around the Ni impurities. These measurements ensure that dissipation in the resistive state of these samples is due to phase slip processes. In addition, these measurements also clearly delineate QPS and TPS dominated temperature regions in these nanowires.

\subsection{Magnetoresistance in the resistive state of NiBi$_3$ nanowires}
In this section we discuss the low field magnetoresistance of the high-Ni and low-Ni samples as a function of temperature. Measurements were performed with a current of 70 $\mu$A within a field range of $\sim$ 2800 Oe between 1.8 K and 5 K. The present work is concerned with the resistive state of the nanowire, where the current density is important in the context of the phase slip. With increasing current density, more phase slip events occur and may drive the nanowire into a fully normal state\cite{delacour2012quantum}. Therefore MR measurements at higher current densities were ruled out to avoid heating effects. Fig. 3 (a) shows the MR of the high-Ni and low-Ni nanowire at 5K, slightly above the transition temperature of the NiBi$_3$. at this temperature both samples show extremely low MR ($\sim$ 5$\times 10^{-4}$ $\%$ at 2800 Oe ) and very similar field dependence. This implies that the Ni impurity concentration does not play a significant role in determining the normal state MR of the nanowires. Apart from a two orders of magnitude increase in the MR, compared to the normal state, two distinct differences are notable in the MR at
\begin{figure}[htbp!]
	\centering
\includegraphics[width=12cm]{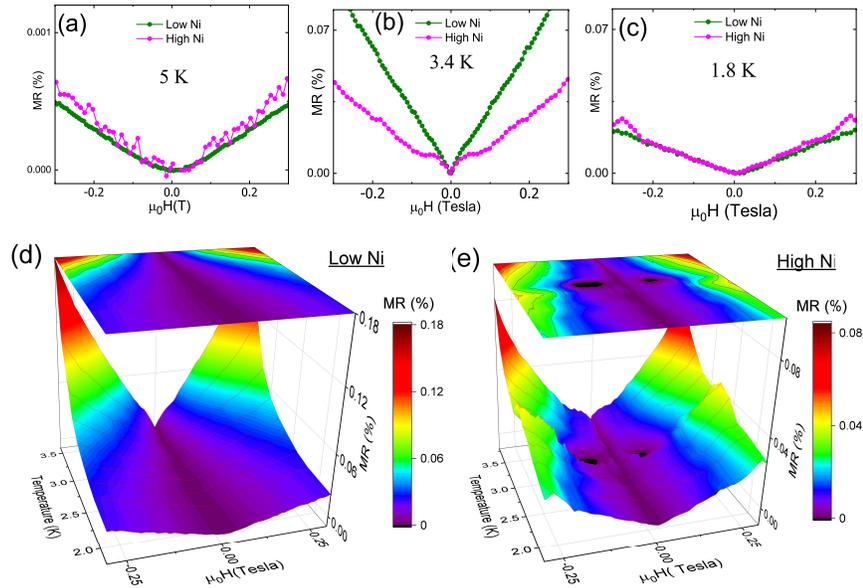}\\
	\caption{Panels (a), (b), and (c) show the comparison of magnetoresistance for low-Ni and high-Ni NiBi$_3$ nanowires at 5K, 3.4 K and 1.8K, respectively. In the normal state of NiBi$_3$ at 5 K, both nanowires have same value and form of MR. Both wires have the same MR at 1.8 K too. However, at 3.4 K the MR of the high-Ni sample shows oscillatory behavior unlike the low-Ni sample. Oscillations at several temperatures, after subtracting a smooth polynomial background, are shown in the supplementary Fig. S2. The 3D surface map and the corresponding contour maps for the MR measurements performed at various temperature are shown in the panels (d) and (e) for the low-Ni and high-Ni nanowires. The high-Ni sample shows a clear negative MR region, marked black in the color map in panel (e), unlike the smooth temperature dependence of MR for the low-Ni nanowire shown in panel (d). }
\end{figure}
3.4 K (Fig 3 (b)). Firstly, the MR of the low-Ni nanowire is almost double of the MR of high-Ni nanowire, which again confirms that the magnetoresistance is not determined by the Ni impurities in NiBi$_3$. Secondly, the high-Ni nanowire shows an oscillatory field dependence of MR, unlike the low-Ni nanowire which shows almost linear field dependence. At the lowest available temperature of 1.8 K, low field MR of both wires almost overlapped with each other (Fig 3 (c)). This shows that the field dependent orbital pair breaking effects are similar in both nanowires. We have plotted the MR data measured at several temperatures as a 3D surface maps in Fig. 3 (d) and (e), for low-Ni and high-Ni nanowires, respectively. The color scale represents the MR percentage in these figures. Corresponding 2D surface projections are also shown in these figures. From these figures it is apparent that while the low-Ni nanowire shows an almost monotonic temperature dependence of MR, the high-Ni sample response is distinctly non-monotonic. In addition to oscillatory field dependence of MR, in a certain temperature range, the high-Ni nanowire also shows negative MR, which corresponds to the black regions in the color map in Fig. 3 (e).

\section{Discussion}
An early signature of negative MR (nMR) in superconducting Pb stripes\cite{Kadin} has been attributed to non-equilibrium charge imbalance effect at normal superconductor interface. Typically, a diffuse normal-superconductor interface forms around phase slip centers where Andreev reflection process introduces charge imbalance over a length scale of $\lambda_Q$, inversely proportional to the normal state resistivity of the materials\cite{Schmid}. It has been shown that a small external magnetic field decreases\cite{Schmid} $\lambda_Q$. Consequently, the resistance of the phase slip process decreases and an overall drop in the resistance of nanowires is possible as a function of magnetic field\cite{vodolazov}. More recent observations of large negative MR in NbN nanowires\cite{Lior,NbNNeg} and a:InO nanowires\cite{Mitra} have also been interpreted in the same manner. The explicit field dependence of the resistance of the phase slip center\cite{Schmid,vodolazov}, however, does not account for a non-monotonic temperature dependence of this effect. Therefore, this mechanism can be ruled out in the context of our data. Another mechanism for nMR has been proposed by Rogachev et al.\cite{Rogachev}, in Nb and MoGe nanowires. 
 \begin{figure}[htbp!]
	\centering
	\includegraphics[width=12cm]{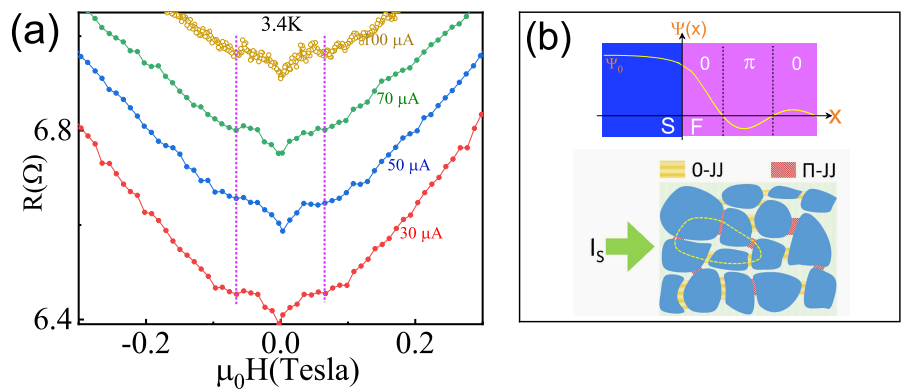}\\
	\caption{(a) Current dependence of magnetoresistance for high-Ni nanowire measured at 3.4 K. MR oscillations do not seem to depend on the bias current, ruling out the possibility of flux pinning related MR oscillations, as discussed in the text. (b) The top inset shows the schematic variation of the amplitude of the superconducting order parameter inside a ferromagnetic layer near a superconductor. Another superconductor can couple to the first superconducting layer in zero or $\pi$ mode depending on their range of separation. The bottom inset shows a schematic internal granular structure of the current carrying nanowire, with random "zero" and "$\pi$" coupled grains. The dashed line shows an example of a possible current loop formed by positive and negative supercurrents across gains. Flux quantization through such loops can lead to oscillatory MR in superconducting nanowires. 
	}
\end{figure}
They have conjectured that stray magnetic moments may form on the nanowire, whose pair breaking effects are suppressed at higher fields causing a negative MR in the nanowires. In such cases, nMR effect must start from zero field itself. Although Rogachev et al.\cite{Rogachev} have not identified any particular magnetic impurities in their nanowires, the NiBi$_3$ nanowires in the present case certainly has Ni impurities. We can, however, rule out this possibility because, (i) nMR in Fig 3 (e) does not start at zero field, and (ii) the negative MR effect appears only in a certain temperature range. Negative MR reported by Tian et al.\cite{Tian} in Zn nanowires was explained by Fu et al.\cite{Fu} as a dissipative boundary effect of the contact pads, where field induced suppression of superconductivity in the contact pads allowed dissipation of vortex-antivortex pairs from the nanowire and promoted nMR. Negative MR in suspended NbN nanowires\cite{NbNNeg} have also been interpreted in the same way. This effect, however, shows a very sharp field dependence\cite{Tian} unlike the present case. Chen et al.\cite{ChenGoldmann} have speculated that reduced phase fluctuations due to field driven enhancement of the dissipative quasiparticle channel in nanowires may have been the origin of negative MR in their Zn nanowires. A monotonic temperature dependence of nMR is expected in such cases, unlike our data. Xiong et al.\cite{Xiong} have reported monotonic temperature dependence of nMR in superconducting Pb nanowires very close to the transition regime. They have speculated that fluctuations in the sign and amplitude of superconducting order parameter may be the cause of this effect. In these experiments, however, there is no clear basis for the negative sign of the order parameter. In contrast, the magnetic inter grain regions in our NiBi$_3$ nanowire is a clear precursor for realizing negative inter grain currents. 

The other important aspect of Fig. 3 (b) is the oscillations in the magnetoresistance for the high-Ni nanowire. Although faint, two symmatrical oscillations are identifiable after subtracting a smooth polynomial baseline(Supplementary Fig S2). Incoherent penetration of vortex bundles into the nanowires at various spots of suppressed edge barriers can cause MR oscillations\cite{Brongersma} via flux matching effects\cite{fluxmatching}. In that scenario, the period of the oscillations should be a function of magnetic field, unlike our data (Fig. 3). A matching effects can also arise from commensurable pinning force density with the density of vortices\cite{fluxmatching1,fluxmatching2,gurevich}, resulting in periodic MR. However, Gurevich et al.\cite{gurevich} have shown that this effect is strongly current dependent. In Fig 4 (a) we show the MR of the high-Ni nanowire, measured at 3.4K at different currents. The period as well as the depth of the resistance minima remained almost unchanged with current. Therefore, matching effect is very unlikely in this case. Another class of interpretations for the oscillatory MR in nanowires is the flux quantization effects through some effective phase-coherent loop formed by the current distribution in the nanowire. MR oscillations in granular Sn nanowires\cite{Herzog}, granular NbN nanowires\cite{patel}, and suspended a:InO nanowires\cite{johansson} have been interpreted in this way. A similar effect may be the origin of MR oscillations in high-Ni NiBi$_3$. 

The only unified model describing both nMR and oscillatory MR in a superconductor was proposed by Kivelson and Spivak\cite{Kivelson}. The authors have shown that large fluctuations in amplitude and sign of the order parameter in a superconducting systems can naturally lead to both these effects. Although fluctuations in the amplitude of the order parameter can be expected in granular SC, no clear scope for a fluctuation in the sign of the order parameter has been identified in the earlier reported nanowire systems\cite{Lior,NbNNeg,Xiong,Herzog}. The NiBi$_3$ nanowire system is unique in this respect as the magnetic inter-grain regions provide ample scope for both sign and magnitude fluctuations of order parameter. When two superconducting regions are separated by a ferromagnetic region, it is now well established that the coupling between the two superconductors can be either 0-type or $\pi$-type, depending on the actual spatial separation between the two\cite{Buzdin,Robinson_Ni}, as shown in the first schematic in the Fig 4(b). In the case of the Nickel barrier (inter-grain region, in this case) it has been shown\cite{Robinson_Ni} that a 0-coupling can change to $\pi$-type coupling and vise-versa, within a length scale of 3-4 nm. Therefore, in this case, it is reasonable to assume that the intergrain Josephson coupling can have a random distribution of 0-type and $\pi$-type couplings. A schematic random couplings between the superconducting grains in the nanowire shown in the Fig 4(b). The observation of MR oscillations is possible only when some effective flux quantization loops (supercurrent loops) form in the nanowire. Since the external current is biased to flow only in one direction, (as shown in the schematic in the Fig 4(b)), effective supercurrent loops are possible only when there are some $\pi$-coupled grains in the sample which carry a local negative critical current, opposite to the external bias current. The dotted line shown in the schematic in Fig 4(b) shows an example of Such a possible quantization loop. The oscillation periods would then be determined by the average loop size due to the distribution of inter grain separations.

\section{Conclusions}
In summary, NiBi$_3$ nanowire is the only system which allows fluctuations in both magnitude and sign of the order parameter, due to the natural granularity and magnetic inter grain regions, respectively. Magnetoresistance of a NiBi$_3$ nanowire with high Ni impurity concentration was compared with the MR of a nanowire containing negligible amount of Ni impurity in their phase-slip driven resistive states, below the superconducting transitions. It was found that the resistance of the nanowire with high Ni impurity undergoes periodic oscillations as a function of magnetic field. In addition, a negative MR also evolves in certain temperature range. Both these effects were absent in the NiBi$_3$ nanowore containing negligible Ni impurities. Observation of both effects (oscillatory MR and nMR) in a single system strongly supports the model of Kivelson $\&$ Spivak, which shows that both these effects are natural outcomes of fluctuations in amplitude and sign of the superconducting order parameter in a system. As superconducting nanowires are highly desirable as lossless interconnects for nanoscale quantum devices \cite{likharev2022dynamics,mooij2006superconducting}, study of macroscopic quantum phenomena in nanowire systems is important. In this respect, NiBi$_3$ is an unique quantum system to study in more detail.

\section*{Acknowledgements}
We acknowledge the funding from National Institute of Science Education and Research (NISER), HBNI, Department of atomic energy (DAE), India, through plan project "RIN-4001".

\section*{References}
\bibliographystyle{iopart-num}
\bibliography{NiBi3_MR_SUST}

\newpage
\section*{Supplementary Information}

\subsection*{I. High resolution TEM}
High resolution TEM imaging was performed on several flakes of the NiBi$_3$ films dispersed on a copper grid. A representative image is shown in the supplementary Fig S1. A Jeol model F200 microscope was used at 200keV for imaging the sample. The inter planar spacing of visible grains were calculated by averaging over 10 successive fringe patterns via the ImageJ software. Corresponding to all the identified d-values of the grains, X-ray diffraction peaks are present in the Fig. 1 (d).    
\begin{figure}[htbp!]
	\centering
	\includegraphics[width=12cm]{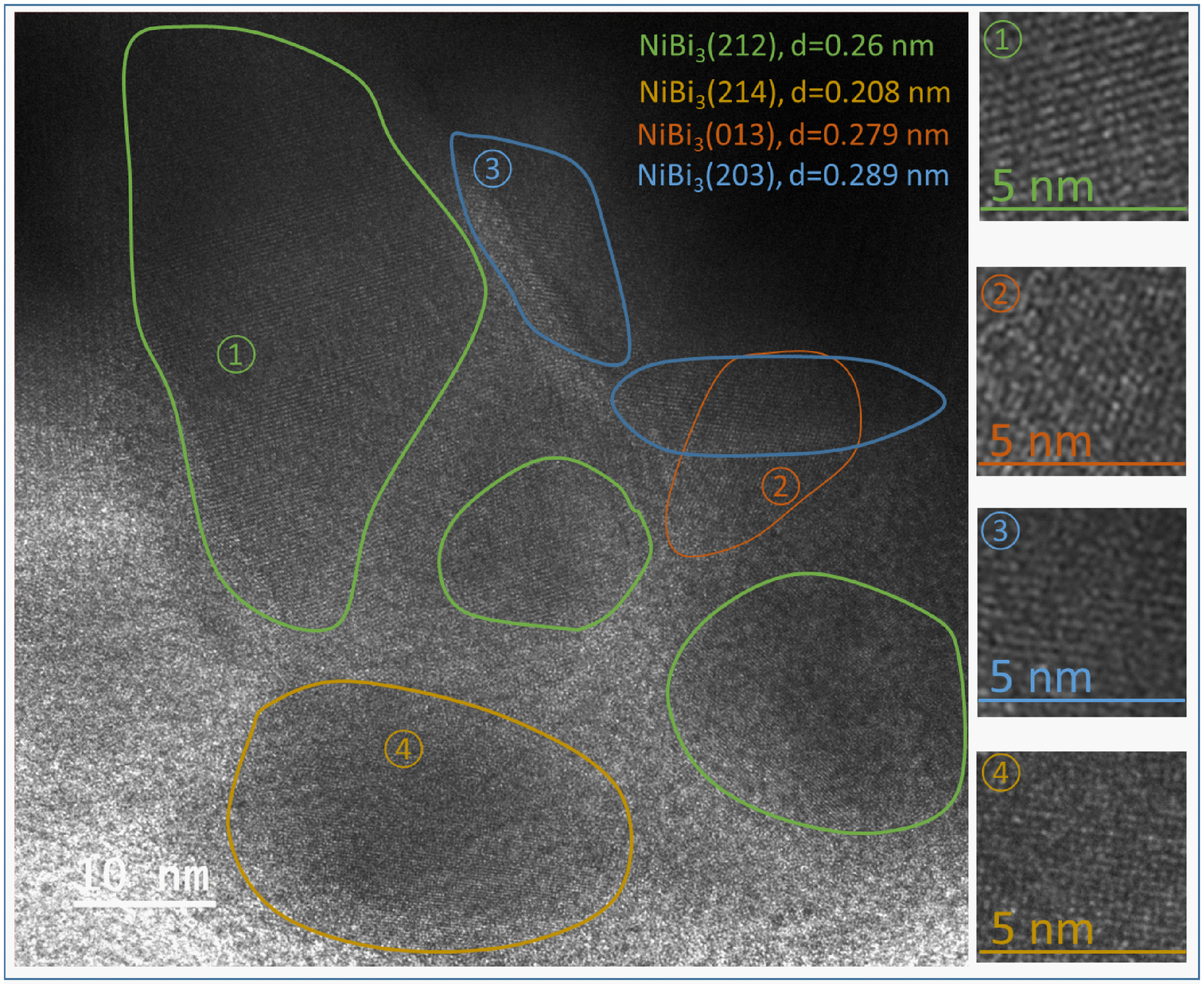}\\
 {\bf Figure S1:} HRTEM images of the NiBi$-3$ film at 200 keV. The identified grains of different orientations have been marked in different colors. The magnified view of the fringe pattern from the identified grains are also shown on the side panels.   
	
\end{figure}
\subsection*{II. Fitting the temperature dependence of resistance of the nanowires at low temperatures}
 The LAMH theory ascribes the thermally activated phase slips near the transition temperature to a resistance which scales as exp(-$\Delta F$/kT), where $\Delta F$ is the superconducting condensate energy barrier. At lower temperatures, especially below $\sim$0.7T$_C$, quantum phase slips also contribute to the resistance of the nanowires. This has been explained by Giordano \cite{giodano5} as a macroscopic quantum tunnelling process through a tilted washboard potential. In this model, the effective resistance at the lowest temperatures increases with increasing bias currents due to the change in slope of the tilted washboard potential\cite{giodano5} by the relation, 
 \begin{equation}
\Delta F_{I}=\pm h I / 2 e
\end{equation}
 where $\Delta F_{I}$ is the difference in free energy between two adjacent minima, e is the electronic charge, I is the current. 
 
 Zero-field resistance vs temperature curves were measured at different bias currents for the low-Ni and high-Ni nanowires. With increasing bias current the probability of phase slips also increases\cite{delacour2012quantum,zhao2016quantum} resulting in a higher resistance of the resistive state of the nanowire. The measured resistances curves below 0.7T$_c$ were fitted as the sum of two terms due to TAPS and QPS \cite{bezryadin2008quantum,zhao2016quantum} following   
 
\begin{equation}
\begin{aligned}
R &=R_{\mathrm{TAPS}}+R_{\mathrm{QPS}} \\
&=\frac{\pi \hbar^{2} \Omega_{\mathrm{TAPS}}}{2 e^{2} k T} \exp \left(-\frac{\Delta F_{\mathrm{TAPS}}}{k T}\right)
+\frac{\pi \hbar^{2} \Omega_{\mathrm{QPS}}}{2 e^{2}\left(\frac{\hbar}{\tau_{\mathrm{GL}}}\right)} \exp \left(-\frac{\Delta F_{\mathrm{QPS}} \tau_{\mathrm{GL}}}{\hbar}\right).
\end{aligned}
\end{equation}
The first and second term of the right hand side of Eqn. 2 corresponds to the thermally activated phase slip resistance and quantum phase slip resistance, respectively. \\
Here, $\Omega_{TAPS}$= (L/$\xi$)($\Delta$F$_{TAPS(QPS)}$/kT)$^{1/2}$ is the attempt frequency\cite{lau2001quantum}, $\Delta$F$_{TAPS}$ and $\Delta$F$_{QPS}$ are TAPS and QPS energy barriers, $\tau_{GL}$= $\pi\hbar$/8k(T$_c$-T) is the Ginzburg-Landau relaxation time which is also considered as tunneling time of a QPS\cite{lau2001quantum}.

The Eqn. 2, rewritten in terms of normalized resistance by r=R/R$_N$ and normalized temperature t=T/T$_c$\cite{zhao2016quantum,bezryadin2008quantum} is, 
\begin{equation}
\begin{aligned}
r=& \frac{R}{R_{N}}=a t^{-3 / 2}(1-t)^{9 / 4} \exp \left(\frac{-b(1-t)^{3 / 2}}{t}\right) \\
&+c(1-t)^{3 / 4} \exp \left(-d(1-t)^{1 / 2}\right)
\end{aligned}
\end{equation}
We fit our RT data according to the Eqn. 3 where a,b,c,d are temperature independent, dimensionless parameters of the model, taken as fitting parameters.  

\subsection*{III. Oscillatory MR in high-Ni NiBi$_3$ nanowires}
The high-Ni NiBi$_3$ nanowire showed oscillation in magneto resistance at all temperatures in the resistive state. The effect was most pronounced around 3.4 K. In order to clearly highlight the periodic oscillations we have subtracted a smooth polynomial background from the MR. In the supplementary Fig S2 we have plotted $\Delta$MR after the subtraction of polynomial baseline at several temperatures. The original MR data along with the polynomial background are also shown in this figure.    

\begin{figure}[htbp!]
	\centering
	\includegraphics[width=14cm]{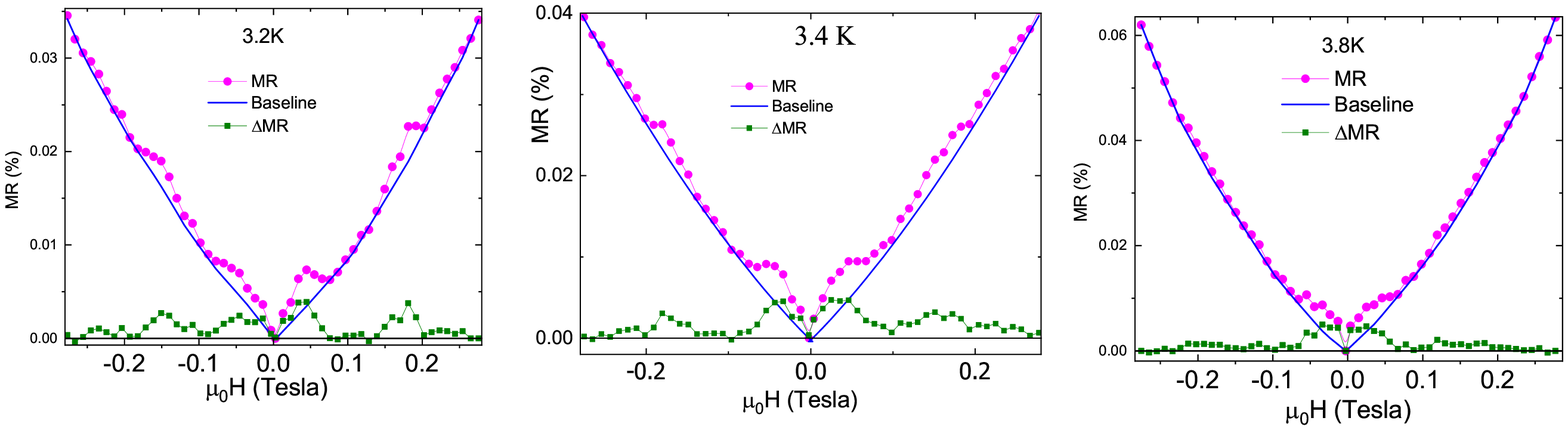}\\
 {\bf Figure S2:} Magneto-resistance as a function of applied magnetic field are shown at 3.2 K, 3.4 K, and 3.8K. The polynomial baseline in each case, subtracted from the respective MR data, are also shown in this figure as solid lines. After subtraction of the baseline the resultant $\Delta$MR shows clear symmetric oscillations.    
	
\end{figure}

\subsection*{IV. Fabrication of NiBi$_3$ films and nanowires}
Thin films of NiBi$_3$ were prepared by co-evaporation of Bi (99.99$\%$) and Ni (99.999$\%$) in a high vacuum chamber with base pressure of 1$\times$10$^{-7}$ mbar. Si wafers with about 300 nm thick thermally grown SiO$_2$ layers were used as substrates, after triple cleaning in Acetone, IPA and water. While e-beam evaporation was used for evaporation of Ni, a  thermal evaporation port was used for evaporation of Bi, due to the low melting point of Bi. We obtained a high transition temperature for the films (close to 4.5 K), even better than the reported T$_c$ of $\sim$ 4.2 K for single crystal NiBi$_3$. The high transition temperature itself testifies to the fact that the superconducting grains are pure and free from Ni impurities, which otherwise would have suppressed the transition temperature significantly. X-ray diffraction measurements (Fig 1(d)), however, shows the presence of Ni (111) peak which implies that some unreacted Ni remains in the inter-grain regions. Therefore, the inter grain regions are naturally expected to become magnetic. 
\begin{figure}[htbp!]
	\centering
	\includegraphics[width=12cm]{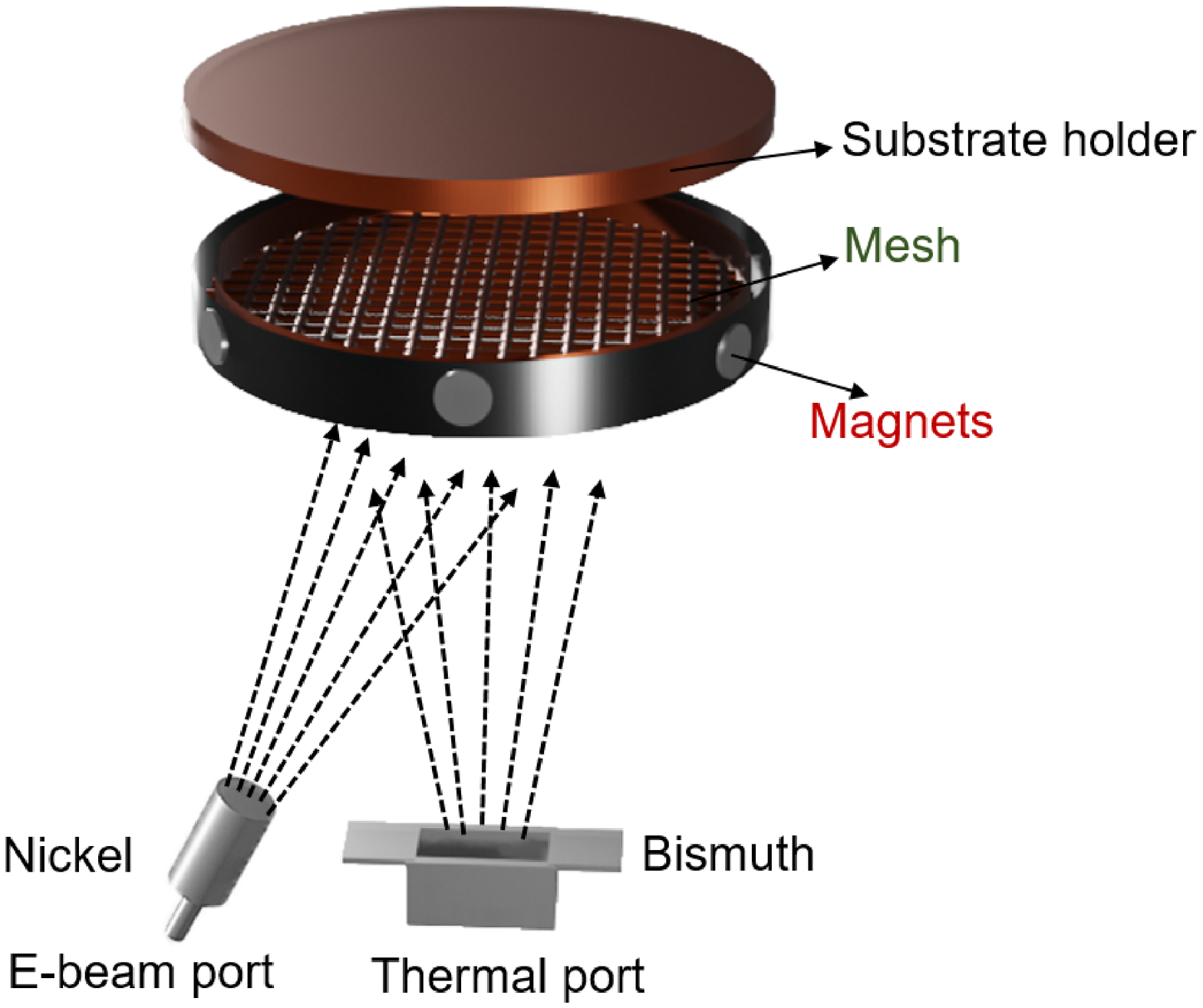}\\
 {\bf Figure S3:} Schematic arrangement of the co-evaporation process. While Ni was e-beam evaporated, Bi was thermally evaporated.
 \end{figure}
 For the purpose of this work we needed another NiBi$_3$ sample in which the inter-grain region is free from Ni. For obtaining this particular sample, we used an iron grid in front of the substrates which was kept magnetized with the help of 6 FeNdB permanent magnets mounted around the periphery of the grid, as shown in the schematic diagram Fig S3. The purpose of this grid was to filter out the unreacted Ni from the physical vapor reaching the substrate, while allowing only the NiBi$_3$ formed in the co-evaporated vapour of Bi and Ni. The X-ray diffraction pattern showed no signature of Ni in the films deposited in this filtering process as shown in the XRD pattern of Fig 1(d) in the main text. EDAX measurements showed that the low-Ni sample has an almost stoichiometric Ni to Bi atomic ratio of $26/74$ while the high-Ni sample has a Ni to Bi ratio of $32/68$ .  

The nanowires of NiBi$_3$ were fabricated from the films only. Since transport measurements require contact pads for connecting current source and voltmeter, we have used a focused ion beam to define the nanowire and the contact pads at the same time, without needing to deposit a separate contact layer. This process avoids unnecessary issues arising from contact resistances. A focused Gallium ion beam at 30 keV from a Zeiss Crossbeam 340 system was used for this purpose. Prior to the patterning, the dose and exposure time was optimized for milling the NiBi$_3$ in a desired area down to the substrate. As shown in the Fig 1(b) of the main text, the Ga ion beam was scanned across the sample to mill out the film along the path of the beam, thereby defining the nanowire and the contact pads and also isolating them from the rest of the film surface. Since the Ga ion beam milling process was done at a pressure of 8$\times$10$_{-7}$ mbar, the milled NiBi$_3$ was efficiently pumped out of the system without much re-deposition, as shown in the Fig 1(c) of the main text. The same process parameters were used for making both the nanowires from the films with and without Ni.

\end{document}